\documentclass[aps,prl,twocolumn,groupedaddress,showpacs]{revtex4} 
\usepackage{amssymb}

\usepackage{graphicx}

\begin{document}

\title{Continuously Tunable Charge in Andreev Quantum Dots}

\author{I.A.\ Sadovskyy$^{a}$, G.B.\ Lesovik$^{a}$, and G.\ Blatter$^{b}$}

\affiliation{$^{a}$L.D.\ Landau Institute for Theoretical Physics RAS,
   119334 Moscow, Russia}

\affiliation{$^{b}$Theoretische Physik, Schafmattstrasse 32, ETH-Zurich, 
   CH-8093 Z\"urich, Switzerland}

\date{\today}

\begin{abstract}
   We show that a quantum dot connected via tunnel barriers to 
   superconducting leads traps a continuously tunable and hence
   fractional charge. The fractional charge on the island is due 
   to particle-hole symmetry breaking and can be tuned via 
   a gate potential acting on the dot or via changes in the phase 
   difference across the island. We determine the groundstate, equilibrium,
   and excitation charges and show how to identify these quantities
   in an experiment.
\end{abstract}


\maketitle

\begin{figure}[b]
  \includegraphics[width=7.9cm]{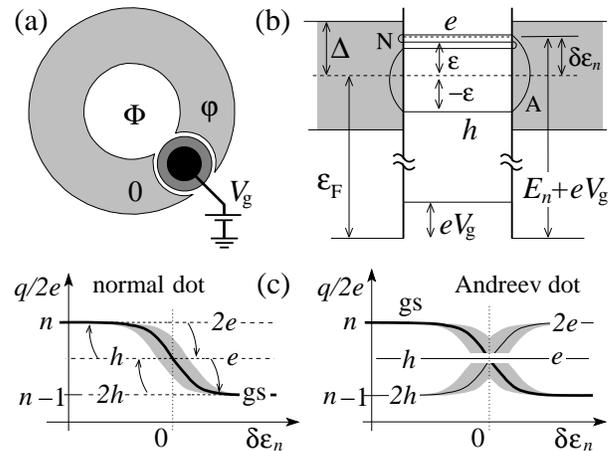}
  \caption[]{(a) Andreev dot (SINIS) embedded in a superconducting
  ring. Tuning of flux $\Phi$ or gate potential $V_\mathrm{g}$ 
  serves to modify the charge on the dot. 
  (b) The mixing of Andreev- and normal scattering at an SIN boundary 
  leads to particle-hole symmetry breaking and to the appearance
  of a fractional charge on the dot. The phenomenon shows up in
  the vicinity of a resonance in the associated NININ device,
  here an electron type resonance, where the electron quasi-particle
  is multiply reflected while the hole component is not.
  (c) Charge trapped on normal- and Andreev dots versus 
  position of resonance $\delta\varepsilon_n$; shown are the 
  ground-state (gs) and single-particle excited-state 
  charges for the singly and doubly occupied situation. gs:
  as the resonance passes the Fermi level, a charge $2e$ is pushed
  out of the dot over a width $\Gamma$; the shaded areas indicate 
  the importance of quantum fluctuations. Dashed lines (normal dot)
  indicate electron- and hole- type resonances; the associated
  trapped charges decay rapidly to the ground state charge (arrows). 
  These resonances are transformed into eigenstates in the Andreev dot;
  they carry integer charge if occupied once (odd-parity) and a 
  fractional charge when occupied twice (even-parity).}
  \label{fig:andreev_dot}
\end{figure}

The superconducting phase $\varphi$ is usually associated with a
dissipation-free flow of current; a phase twist by $2\pi$ within a bulk
superconductor drives a magnetic vortex \cite{Abrikosov} and a phase drop
$\varphi$ across a junction drives a Josephson current \cite{Josephson}. Only
much later it has been recognized that vortices carry a (small) charge as well
\cite{vortexcharge}, an effect which is due to particle-hole symmetry
breaking. In this letter, we show that particle-hole asymmetry also generates
a localized charge in a metallic quantum dot coupled to a superconducting
loop, a so-called Andreev quantum dot \cite{Chtech_Naz_03}, see Fig.\
\ref{fig:andreev_dot}(a). This charge is continuously tunable (and hence
fractional) through the gate potential $V_g$ or via the phase bias $\varphi$
across the dot. It manifests itself for the (even parity) ground- and
doubly-occupied excited states, while the (odd parity) singly-occupied excited
state exhibits an integer charge, see Fig.\ \ref{fig:andreev_dot}(c); as a
corollary, the excitation charges are non-integer as well. The fractional
charge can be observed via measurement of the associated telegraph noise
signal due to the stochastic occupation of the Andreev states or through its
dependence on the flux threading the loop. The latter provides a realization
of a flux-to-charge converter allowing to make use of the device as a
magnetic-flux detector.

A tunable charge in a mesoscopic setup, a ring with a (nearby or interrupting)
quantum dot, has been discussed in Refs.\ \cite{buttiker_stafford,deo}; the
origin of this fractional charge is easily understood in terms of the extended
nature of the wave function, with only a fraction localized on the dot. This
contrasts with the fractional charge discussed here, where the `missing' part
is completely delocalized in the nearby superconducting condensate. Our charge
then reminds about the fractional charge associated with excitations in a
superconductor which has been discussed in the context of charge relaxation
\cite{pethick_smith} (or branch imbalance \cite{blonder_tinkham_klapwijk}) in
non-equilibrium superconductivity; there, too, the missing charge is
`dissolved' in the superconducting condensate. In the following, we first
discuss in detail the origin of the tunable charge which we then determine
quantitatively. We discuss quantum fluctuations in the trapped charge as well
as corrections due to weak Coulomb interactions, see Ref.\
\cite{rozhkov_arovas} for a discussion of strong Coulomb effects and the Kondo
effect in an Andreev dot. We end with a discussion of telegraph noise allowing
to detect the fractional charge.

In superconducting mesoscopic devices the phenomenon of Andreev
scattering \cite{Andreev} plays a central role: electrons (holes)
incident from a normal metal (N) with energies $|E -
\varepsilon_{\rm\scriptscriptstyle F}| < \Delta$ and scattering 
off a superconductor (S) are reflected as holes (electrons) with 
a Cooper-pair entering (leaving) the superconductor (here, 
$\varepsilon_{\rm\scriptscriptstyle F}$ and $\Delta$ denote the
Fermi energy and the superconducting gap). In an SNS junction, 
electrons convert to holes at one NS boundary and back to electrons
at the opposite NS contact, thus generating a particle-hole symmetric
uncharged bound state (we ignore a small electron-hole symmetry 
breaking of order $\Delta/\varepsilon_{\rm\scriptscriptstyle F}$).
When the NS interfaces are augmented by thin insulating layers 
I in an SINIS junction, the particles in the normal-metal region 
undergo imperfect Andreev scattering at the NIS boundaries with 
mixed Andreev- and normal scattering processes; the resulting 
Andreev levels involve unequal electron- and hole-like components 
and hence particle-hole symmetry is broken, see Fig.\ 
\ref{fig:andreev_dot}(b). Alternatively, the Andreev state 
can be understood as deriving from normal resonances in the 
associated NININ structure --- replacing the normal leads by
superconducting ones, a resonance residing in the gap region 
transforms into a sharp Andreev level with predominantly
electron- (resonance above $\varepsilon_{\rm\scriptscriptstyle 
F}$) or hole character (resonance below $\varepsilon_{\rm
\scriptscriptstyle F}$). The properties of these trapped Andreev 
states are easily tuned, either through shifts in their position
within the gap via changing gate bias $V_g$ or via changes in the
phase $\varphi$ across the superconducting banks.

The Andreev states give rise to new opportunities for tunable Josephson
devices, e.g., the Josephson transistor \cite{JT,kuhn_01}; here, we are
interested in their charging properties.  Below, we consider small islands
with a large separation $\delta_n$ of resonances in the associated NININ
problem, $\delta_n > \Delta$, such that a single Andreev level is trapped
within the gap region. Such Andreev dots have recently been fabricated by
coupling carbon nanotubes to superconducting banks \cite{cnt_K,cnt_JT}, with
the main focus on the competition between the Kondo effect and
superconductivity \cite{cnt_K}.  We are interested in sufficiently well
isolated islands with a small width $\Gamma_n$ of the associated NININ
resonance, $\Gamma_n < \Delta$. Also, we wish to neglect charging effects in
our first analysis here and hence demand that the Coulomb energy is small,
$E_\mathrm{C} < \Gamma_n$. In summary, our device operates with energy scales
$E_\mathrm{C} < \Gamma_n < \Delta < \delta_n$, a situation which can be
generated in the lab, see our estimates below. In what follows, we determine
the charge configuration of the ground- and excited states versus
$V_g$ and $\varphi$, find the thermal equilibrium charge and discuss the
excitation dynamics of the Andreev states (due to elastic and Coulomb
interactions) providing the characteristics of the telegraph noise in the
charge state of the dot.

The resonances in the NININ setup derive from the eigenvalue problem $H_0
\Psi=E\Psi$ with $H_0=-\hbar^2\partial_x^2/2m +U(x)-\varepsilon_{\rm
\scriptscriptstyle F}$ and the potential $U(x)$ $=U_\mathrm{ps}(x+L/2)
+U_\mathrm{ps}(x-L/2)]+eV_g\Theta(L/2-|x|)]$ describing two point-scatterers
(with transmission and reflection coefficients $\sqrt{T}e^{i\chi^t}$ and
$\sqrt{R}e^{i\chi^r}$) and the effect of the gate potential $V_g$, which we
assume to be small as compared to the particle's energy $E$ (measured from the
band bottom in the leads), $eV_g \ll E$.  Resonances then appear at energies
$E_n = \varepsilon_L (n\pi-\chi^r)^2$; they are separated by $\delta_n
=(E_{n+1}-E_{n-1})/2 \approx 2E_n/n$ and are characterized by the width
$\Gamma_n = T\delta_n/\pi\sqrt{R}$, where $\varepsilon_L=\hbar^2/2mL^2$. The
bias $V_g$ shifts the resonances by $eV_g$; we denote the position of the
$n$-th resonance relative to $\varepsilon_{\rm\scriptscriptstyle F}$ by
$\delta\varepsilon_n = E_n+eV_g-\varepsilon_{\rm \scriptscriptstyle F}$, see
Fig.\ \ref{fig:andreev_dot}(b).

We go from a normal- to an Andreev dot by replacing the normal leads 
with superconducting ones. In order to include Andreev scattering 
in the SINIS setup we have to solve the Bogoliubov-de Gennes 
equations (we choose states with $\varepsilon \geq 0$, see Fig.\ 
\ref{fig:andreev_dot}(b))
\begin{eqnarray}
\left[ \begin{array}{cc}
  \!H_0 & \!\! \Delta(x) \! \\
  \!\! \Delta^*(x) & \!\!-H_0 \!
  \end{array}\right]
  \! \left[\begin{array}{c}\! u_n(x)\! \\
  \! v_n(x) \! \end{array} \right] = \varepsilon^{\rm\scriptscriptstyle A}_n
  \! \left[\begin{array}{c}\! u_n(x)\! \\
  \! v_n(x) \! \end{array} \right],
\end{eqnarray}
with the pairing potential $\Delta(x)=\Delta[\Theta(-x-L/2) e^{-i\varphi/2}$
$+\Theta(x-L/2) e^{i\varphi/2}]$; $u_n(x)$ and $v_n(x)$ are the electron-
and hole-like components of the wave function. The discrete states
trapped below the gap derive from the quantization condition (in Andreev
approximation; the bar `$\bar{\cdot}$' and indices `$\pm$' refer to scattering
probabilities and phases for the NININ setup and energies $\varepsilon_{\rm 
\scriptscriptstyle F}\pm \varepsilon$)
\begin{eqnarray}
   \label{qc}
   \cos(S_+-S_- -2\alpha)
   =\!\sqrt{\bar{R}_+\bar{R}_-}\cos\beta+\!\sqrt{\bar{T}_+\bar{T}_-}\cos\varphi.
\end{eqnarray}
The phase $\alpha=\arccos(\varepsilon^{\rm\scriptscriptstyle A}_n /\Delta)$
is due to Andreev scattering and the phases $S_\pm = \chi_\pm^t + k_\pm L$,
$k_\pm = \sqrt{2m(\varepsilon_{\rm \scriptscriptstyle F}\pm
\varepsilon)}/\hbar$ account for the propagation across the island; for
symmetric barriers the phase $\beta=(\bar\chi^t_+-\bar\chi^r_+)
-(\bar\chi^t_--\bar\chi^r_-)$ is a multiple of $\pi$ and produces a smooth
function $\sqrt{\bar{R}_+\bar{R}_-} \cos\beta$ changing sign at every
resonance \cite{kuhn_01,clb}.

\begin{figure}[hb]
  \includegraphics[width=7.5cm]{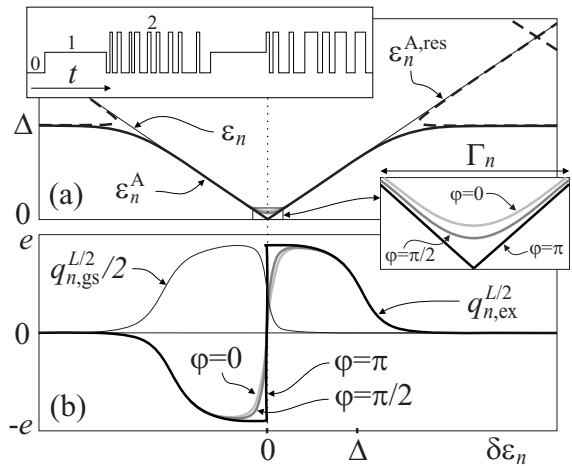}
  \caption[]{(a) Bound state $\varepsilon^{\rm\scriptscriptstyle A}_n$
  (thick solid lines) and resonances $\varepsilon^{\rm\scriptscriptstyle 
  A,res}_n$ (dashed lines) in the Andreev dot versus position $\delta
  \varepsilon_n$ of the normal state resonance (tuned via the gate 
  potential $V_g$); we choose $\Gamma_n = 0.2 \Delta$. Also shown 
  are the normal resonances $\varepsilon_n$ (thin solid line) and 
  the dependence on the phase difference $\varphi$ across the dot (inset). 
  (b) Charges $q^{\scriptscriptstyle L/2}_{n,{\rm ex}}$ and 
  $q^{\scriptscriptstyle L/2}_{n,{\rm gs}}$.
  Top inset: structure of telegraph noise with long periods of
  ground state (0) and singly occupied (1) states and the ground state
  trace interrupted by fast doubly excited states (2).}
  \label{fig:en_q}
\end{figure}

The solution of (\ref{qc}) is straightforward and provides the 
spectrum shown in Fig.\ \ref{fig:en_q}(a). A small dot $L 
\lesssim \xi$ ($\xi$ the superconducting coherence length)
results in a single Andreev state which lives near 
(but below) the gap $\Delta$ when the dot is tuned far off resonance 
with $|\delta \varepsilon_n| \gg \Delta$, $\varepsilon^{\rm 
\scriptscriptstyle A}_n \approx \Delta(1-\Gamma_n^2/8\delta
\varepsilon_n^2)$; when the resonance resides in the gap region, 
$|\delta \varepsilon_n| < \Delta$, the associated Andreev level 
approaches the Fermi energy $\varepsilon_{\rm \scriptscriptstyle F}$ 
linearly,
\begin{equation}
   \varepsilon^{\rm\scriptscriptstyle A}_n \approx (1-\Gamma_n/2\Delta)
   |\delta\varepsilon_n|,
   \label{eA_linear}
\end{equation}
where we have assumed $\Gamma_n \ll \Delta$. As $\varepsilon^{\rm
\scriptscriptstyle A}_n$ drops below the resonance width $\Gamma_n$ 
the Andreev state becomes sensitive to the phase difference $\varphi$ 
across the junction,
\begin{equation}
   \varepsilon^{\rm\scriptscriptstyle A}_n \approx 
   (1-\Gamma_n/2\Delta)
   \sqrt{\delta\varepsilon_n^2 + (\Gamma_n/2)^2
   \cos^2(\varphi/2)}.
   \label{eA_varphi}
\end{equation}
The resonances above $\Delta$ follow closely the normal resonances 
$\varepsilon_n$, see Fig.\ \ref{fig:en_q}(a). 

The Andreev state at $\varepsilon^{\rm\scriptscriptstyle A}_n$ carries
a nontrivial charge which we find as the expectation value of the
charge operator (for the space interval $[-a,a]$) $Q^a\equiv e
\int_{-a}^{a}dx\,\sum_{\sigma=\pm 1}\Psi_\sigma^\dagger(x)\Psi_\sigma(x)$
with $\Psi_\sigma(x)=\sum_n[u_n(x)\alpha_{n,\sigma}+\sigma v_n^*(x) 
\alpha^\dagger_{n,-\sigma}]$ and $\alpha_{n,\sigma},~\alpha^\dagger_{n,\sigma}$ 
the usual Bogoliubov operators; the $\sum_n$ includes localized and 
extended states. Defining the quantities $q_{n,u}^a \equiv 
\int_{-a}^{a} dx\, e|u_n(x)|^2$ and $q_{n,v}^a \equiv \int_{-a}^{a} 
dx\, e|v_n(x)|^2$ (with the normalization $q_{n,u}^\infty + 
q_{n,v}^\infty = e$), we obtain the thermal equilibrium charge 
(we subtract the charge $2e(n-1)$ due to filled resonances and 
continuum states)
\begin{equation}
   q^a_{n,{\rm eq}} = 2 f(\varepsilon^{\rm\scriptscriptstyle A}_n)\,
   q_{n,u}^a+2[1-f(\varepsilon^{\rm\scriptscriptstyle A}_n)]\,q_{n,v}^a;
   \label{qeq}
\end{equation}
setting the Fermi function $f(\varepsilon)$ to zero we obtain the 
ground state charge $q^a_{n,{\rm gs}}\equiv \langle 0|Q^a|0\rangle 
= 2q_{n,v}^a$. The excitation charge (of one spin state $|\sigma\rangle$)
assumes the value 
\begin{equation}
   q^a_{n,{\rm ex}} \equiv 
   \langle \sigma|Q^a|\sigma\rangle-\langle 0|Q^a|0\rangle
    = q_{n,u}^a- q_{n,v}^a.
   \label{qex}
\end{equation}
These charges are localized around the dot, either strictly within 
the region $[-L/2.L/2]$ as for the excitation charge $q^\infty_{n,
{\rm ex}}=q^{\scriptscriptstyle L/2}_{n,{\rm ex}}$ or with an additional 
extension of size $\sim \xi$ around the dot carrying a small weight
\cite{weight}, $q^\infty_{n,{\rm eq}} \approx q^{\scriptscriptstyle 
L/2}_{n,{\rm eq}}$ (hence, the fractional charge is {\it not} due 
to a missing part of a wave function). The excitation charge can 
be directly obtained from the voltage dispersion of the Andreev 
state, $q^{\scriptscriptstyle L/2}_{n,{\rm ex}}=\partial_{V_g}
\varepsilon^{\rm \scriptscriptstyle A}_n$; combining this with the 
normalization condition, the other charges $q^\infty_{n,{\rm eq}}$ 
and $ q^\infty_{n,{\rm gs}}$ follow immediately. The expressions 
(\ref{eA_linear}) and (\ref{eA_varphi}) for the position
$\varepsilon^{\rm\scriptscriptstyle A}_n$ provides the 
excitation charge,
\begin{eqnarray}
   q^{\scriptscriptstyle L/2}_{n,{\rm ex}} \approx 
   \left\{ \begin{array}{r@{\quad}l}
   \displaystyle{\frac{e\, (1-\Gamma_n/2\Delta)
   \delta\varepsilon_n}{\sqrt{\delta\varepsilon_n^2+(\Gamma_n/2)^2
   \cos^2(\varphi/2)}},}
   & {\delta\varepsilon_n < \Gamma_n,}\\ \noalign{\vskip 5 pt}
   \displaystyle{\mathrm{sign}(\delta\varepsilon_n)
   \, e\, (1-\Gamma_n/2\Delta),}
   & {\delta\varepsilon_n < \Delta,}\\ \noalign{\vskip 5 pt}
   \displaystyle{\mathrm{sign}(\delta\varepsilon_n)
   {e\, \Delta\Gamma_n^2}/{4\delta\varepsilon_n^3},}
   & {\Delta < \delta\varepsilon_n.}
   \end{array}\right.
   \label{qex_analyt}
\end{eqnarray}
The full numerical result is shown in Fig.\ \ref{fig:en_q}(b):
the charge rises linearly (with slope $\approx 2e/\Gamma_n 
\cos(\varphi/2)$; note the sharp change in charge for a 
$\pi$-tuned Andreev dot at $\varphi = \pi$) as the normal 
resonance crosses the Fermi level and then saturates at a value 
$e\,(1-\Gamma_n/2\Delta)$. As the normal resonance leaves the
gap region, $|\delta\varepsilon_n| >\Delta$, the charge 
$q^{\scriptscriptstyle L/2}_{n, {\rm ex}}$ vanishes $\propto 
\Delta \Gamma_n^2 / \delta\varepsilon_n^3$. 
We see, that tunable fractional and localized ground state- and 
excitation charges appear on the Andreev dot each time a normal 
resonance crosses the Fermi level; this is the main result of our 
paper. Note that the addition of the (fractional) ground-state-
and excitation charges conspires to produce the integer charge 
of the singly-occupied odd-parity excited state, $\langle \sigma|
Q^{\scriptscriptstyle L/2}|\sigma\rangle=q_{n,u}^{\scriptscriptstyle 
L/2} + q_{n,v}^{\scriptscriptstyle L/2} \approx e$.

The above non-trivial charges are a consequence of the broken 
particle-hole symmetry and result from a superposition of 
states with definite integer charge, electron- and hole-like. 
As a result, the charges will undergo quantum fluctuations 
which are quantified by the variance $\delta q^a \equiv [\langle 
(Q^a)^2\rangle - \langle Q^a \rangle^2]^{1/2}$. Here, we are 
interested in the charge fluctuations of the ground- and singly/doubly
excited states $|0\rangle$, $|\pm 1\rangle$, and $|2\rangle$;
with $\langle \nu|(Q^a)^2|\nu\rangle= \sum_\mu |\langle\nu|Q^a|
\mu\rangle|^2$ the determination of $\delta q^a$ involves the 
calculation of matrix elements $\langle\nu|Q^a|\mu\rangle$. 
A detector will measure the charge over some time interval 
$\tau$ such that the relevant charge is given by 
the time average $\bar{Q}^a \equiv \int_0^\tau (dt/\tau) Q^a(t)$; 
as a result, only matrix elements connecting states with 
energy difference less than $\hbar/\tau$ have to be considered;
assuming typical measurement frequencies $1/\tau \ll \Delta/\hbar$,
we can restrict the sum over intermediate states $|\mu\rangle$ 
to the four states $|0\rangle,~|\pm 1\rangle,~|2\rangle$.
The only relevant matrix element then is $\langle 0|Q^a|2\rangle 
= 2e\int_{-a}^{a} dx \, u_n(x) v_n(x)$ which is of order $e$.
Hence, the charges of the singly-occupied states $|\pm 1\rangle$ 
do not fluctuate, while the charges of the ground- and 
doubly-excited states fluctuate strongly, see Fig.\ 
\ref{fig:andreev_dot}(c). 

Next, let us comment on the effects of Coulomb interactions. 
Two important issues are the screening of the dot's charge 
and the mixing between charge states. Screening due to 
additional charge-flow to and from the Andreev dot is 
governed by the density of states. The relevant energy 
scale is given by the distance $\delta_n$ between resonances 
(as these carry large local charges) and hence on-dot charge 
screening is irrelevant if $E_\mathrm{C} \ll \delta_n$. 
Mixing of states is again governed by the matrix elements 
$\langle\nu|(Q^a)^2|\nu\rangle$; restricting the analysis 
to the four states $|0\rangle,~|\pm 1\rangle,~|2\rangle$, 
we find that mixing occurs between the ground- and doubly-excited 
states but is small if $E_\mathrm{C} \ll \varepsilon_n^{\rm
\scriptscriptstyle A}$. The energy scale $E_\mathrm{C} \approx 
e^2/2C$ can be estimated via the capacitance $C \approx \epsilon 
L$ of the Andreev dot with $\epsilon$ the dielectric constant; 
on the other hand, the distance between resonances $\delta_n 
= h v_{\rm \scriptscriptstyle F}/2L$ depends linearly on $L$, 
too. The crucial dimensionless device parameter is the ratio 
$\delta_n/E_\mathrm{C} = h v_{\rm \scriptscriptstyle F} 
\epsilon/e^2$; assuming typical values $\epsilon \sim 10$ 
and $v_{\rm \scriptscriptstyle F} \sim 10^6$ m/s, we obtain 
a ratio $\delta_n/E_\mathrm{C} \approx 30$, allowing for a 
sequence of energies $E_\mathrm{C} < \Gamma_n < \Delta < 
\delta_n$ as required in our setup. In Ref.\ \cite{cnt_JT} an 
Andreev dot in the form of a nanotube with $L \approx 500$ 
nm has been studied, that corresponds to a charging energy 
$E_\mathrm{C}$ of order one Kelvin, hence the above 
inequalities can be satisfied in a realistic device.

The fractional charge of excitations can be detected by various 
means; here, we discuss its observation through the measurement 
of the telegraph noise which arises due to the thermal occupation 
of states (we denote temperature by $\theta$ and set $k_{\rm
\scriptscriptstyle B} =1$). The noise pattern involves the 
three states $|\pm 1\rangle$ and $|2\rangle$; these states 
are filled and emptied through phonon absorption and emission 
processes \cite{fi}; in addition, fluctuations of the gate potential 
provide a competing channel. Two types of processes contribute 
to the telegraph noise, those coupling the ground- and 
singly-occupied states with rates $\gamma_{01}$ and $\gamma_{10} 
= \gamma_{01}\exp(\varepsilon /\theta)$, and those coupling the 
ground- and doubly-occupied states with rates $\gamma_{02}$ 
and $\gamma_{20} =\gamma_{02}\exp(2\varepsilon/\theta)$ 
(we introduce the shorthand $\varepsilon = \varepsilon_n^{\rm 
\scriptscriptstyle A}$; the rates connecting $|\pm 1\rangle$
and $|2\rangle$ agree with the single-particle rates, $\gamma_{01} 
= \gamma_{12}$ and $\gamma_{10} = \gamma_{21}$). The transitions 
$0 \leftrightarrow 1$ involve a second quasiparticle in the 
continuum with energy $E > \Delta$. 

The rates $\gamma_{01}$ and $\gamma_{02}$ are derived from Fermi's 
Golden Rule with the perturbation given by the electron-phonon 
interaction $H = g \int d x \,n_e (\partial_x u)$ with $u$ the 
displacement, $n_e$ the electronic density, and $g$ the 
electron-phonon coupling assuming values of order 1 eV 
(we consider one-dimensional modes for both electrons and 
phonons); an alternative channel is provided by the fluctuating
gate potential with $H = \int d x \,e\, n_e V_g$. 
Concentrating on elastic modes, the determination of 
$\gamma_{02}$ is straightforward and a simple estimate provides 
the result ($N_{2\varepsilon}$ is the Bose factor for the phonon 
state, $a$ and $k>1/L$ denote the lattice constant and the phonon 
wave number) $\gamma_{02} \sim (g^2/\hbar m v^2_{\rm
\scriptscriptstyle F}) (a/kL^2)N_{2\varepsilon}$.
As $k < 1/L$, we enter the zero-dimensional case and the result
crosses over to $\gamma_{02} \sim (g^2/\hbar m v^2_{\rm 
\scriptscriptstyle F})\,ak\, N_{2\varepsilon}$. On the other hand, 
fluctuations of the gate potential contribute the rate 
$\tilde\gamma_{02} \sim (e^2/\hbar C) N_{2\varepsilon}$. 
The numerical estimate with $g\sim 1$ eV, $m v^2_{\rm
\scriptscriptstyle F} \sim 1$ eV, and $L \approx 500$ nm 
provides the result $\gamma_{02}\sim 10^{12}$ s$^{-1} 
N_{2\varepsilon}$ at $k\sim 1/L$, which has to be 
compared with the Coulomb term $\tilde\gamma_{02} \sim 10^{11} 
N_{2\varepsilon}$ s$^{-1}$.

The calculation of the rate $\gamma_{01}$ involves an additional 
sum over continuum states which is dominated by energies 
$E \sim \Delta$; here, we estimate the rate for the simultaneous 
occupation of the Andreev state and a continuum state under 
phonon absorption and find $\gamma_{01} \sim (g^2 T/\hbar m 
v^2_{\rm\scriptscriptstyle F}) (a s/L v_{\rm\scriptscriptstyle F}) 
\sqrt{\theta/\Delta} \,e^{-\Delta/\theta}[1+e^{-\varepsilon/\theta}]$,
where $s$ denotes the speed of sound. Note that $kL \sim \Delta L/\hbar s
\gg 1$ in the present case. Fluctuations of the gate potential contribute
with a rate $\tilde\gamma_{01} \sim\hbar^{-1}(e^2/C)\sqrt{\theta/\Delta}
e^{-\Delta/\theta}[1+e^{-\varepsilon/\theta}]$; using typical parameters
$\theta \approx 0.1 \Delta \approx 1$ K and $v_{\rm\scriptscriptstyle F}/s
\sim 10^3$, we obtain numerical results of order $\gamma_{01} \sim 10^{10} 
{\rm s}^{-1} T \exp(-\Delta/\theta)$ and $\tilde\gamma_{01} \sim 10^{11} 
{\rm s}^{-1} \exp(-\Delta/\theta)$ due to elastic and Coulomb 
interactions, respectively \cite{3D}. Summarizing, we find that 
the processes $0\leftrightarrow 2$ are always fast, 
while the $0\leftrightarrow 1$ transitions involve 
the exponential factor $\exp(-\Delta/\theta)$, allowing us to reduce the 
rate $\gamma_{01}$ dramatically by lowering the temperature. 
Hence, the $0\leftrightarrow 1$ processes can be effectively 
separated from the $0\leftrightarrow 2$ transitions, an effect 
which can be exploited in the experimental detection of the 
fractional charge.

Given today's performance of single electron transistors (SET) 
operating with a sensitivity of $\sim 10^{-5} e/\sqrt{\mathrm{Hz}}$ 
at frequencies $f < 10^9~\mathrm{Hz}$ \cite{aassime_01}, the 
telegraph noise due to the thermal population of the Andreev 
state may be observable and thus the fractional charge on 
the Andreev dot may be measured. However, the processes 
$0 \leftrightarrow 2$ may well be too fast, beyond today's 
time resolution of a SET. In this case we suggest two 
alternative schemes for the measurement of the excitation 
charge: i) Measuring the time averaged equilibrium 
charge $q^a_{n,{\rm eq}}$ once at high temperatures, 
$q^a_{n,{\rm eq}}(\theta \gg \varepsilon)=q^a_{n,u}+q^a_{n,v}$, 
and another time at low temperatures, $q^a_{n,{\rm eq}} (\theta 
\ll \varepsilon) = 2 q^a_{n,v}$, their difference provides the 
excitation charge $q^a_{n,{\rm ex}}=q^a_{n,u}-q^a_{n,v}$. 
ii) A measurement at low temperatures resolving the slow 
$0 \leftrightarrow 1$ transitions and averaging over the 
fast $0 \leftrightarrow 2$ processes, see Fig.\ 
\ref{fig:en_q}, measures the following (average) charges: 
given the occupation probabilities $p_0=(1-f_\varepsilon)^2$, 
$p_1=2 f_\varepsilon(1-f_\varepsilon)$, and $p_2=f_\varepsilon^2$, 
the average over the rapidly fluctuating regimes involving the 
states 0 and 2 provides the charge $\langle q_n^a \rangle_{0,2}
\approx 2q^a_{n,v}+2f_\varepsilon^2 q^a_{n,{\rm ex}}$, while the 
singly charged regime appears with the trivial average $\langle 
q_n^a \rangle_{1} = q^a_{n,v}+q^a_{n,u}$. At low temperatures we 
have $f_\varepsilon \ll 1$ and the difference $\langle q_n^a 
\rangle_{1} - \langle q_n^a \rangle_{0,2} \approx q^a_{n,u}
-q^a_{n,v}$ gives $q^a_{n,{\rm ex}}$. 

In conclusion, we have discussed the peculiar properties of the
trapped charge on an Andreev dot. We have found continuously 
tunable fractional and fluctuating charges for the ground- and 
doubly-excited states, while the odd-parity singly-occupied state 
exhibits an integer charge. The transition dynamics of these 
states allows to observe these charges in the telegraph noise.

We acknowledge financial support by the CTS-ETHZ, the Swiss NSF
through MaNEP, and the Russian Foundation for Basic Research 
(06-02-17086-a; IAS and GBL).


\begin{thebibliography}{99}
\bibitem{weight} The quasiparticles spend a long time
$\tau_{\rm dot} \sim \hbar/\Gamma_n$ in the dot and only a short
time $\tau_{\rm sc} \sim \xi/v_{\rm\scriptscriptstyle F}
\sim \hbar/\Delta$ in the superconductor outside. Note that
$q^\infty_{n,\rm eq}$ starts to differ from $q^{\scriptscriptstyle 
L/2}_{n,\rm eq}$ when $\delta\varepsilon_n > \Delta$, as the charge
residing in the superconductor associated with $\varepsilon_n^{\rm
\scriptscriptstyle A} \approx \Delta$ is no longer small.
\bibitem{3D} Three-dimensional phonon modes lead to rates 
suppressed by large factors $(\varepsilon/\hbar \omega_{\rm
\scriptscriptstyle D})^3 (L/a) \sim 10^{-6}$ and $(\Delta
/\hbar\omega_{\rm\scriptscriptstyle D})^3(L/a) \sim 10^{-3}$, with
$\omega_{\rm \scriptscriptstyle D}$ the Debye frequency and
we assume $\varepsilon/\hbar\omega_{\rm\scriptscriptstyle D} 
\sim 10^{-3}$.
\end{thebibliography}
\end{document}